\documentclass{webofc}
\usepackage[varg]{txfonts}   
\usepackage{color}
\usepackage{natbib}
\usepackage{hyperref}
\definecolor{airforceblue}{rgb}{0.36, 0.54, 0.66}
\definecolor{darkpurple}{rgb}{0.5, 0.2, 0.8}
\definecolor{dartmouthgreen}{rgb}{0.05, 0.5, 0.06}

\begin{document}
\title{Decoding Photons: Physics in the Latent Space of a \mbox{BIB-AE} Generative Network}
%
%

\author{
\firstname{Erik} \lastname{Buhmann}\inst{1}\fnsep\thanks{\email{erik.buhmann@uni-hamburg.de}} \and
\firstname{Sascha} \lastname{Diefenbacher}\inst{1,2} \and
\firstname{Engin} \lastname{Eren}\inst{2} \and
\firstname{Frank} \lastname{Gaede}\inst{2} \and
\firstname{Gregor} \lastname{Kasieczka}\inst{1} \and
\firstname{Anatolii} \lastname{Korol}\inst{3} \and
\firstname{Katja} \lastname{Kr\"uger}\inst{2}
}

\institute{Institut f\"ur Experimentalphysik, Universit\"at Hamburg, Germany
\and
Deutsches Elektronen-Synchrotron, Germany
\and
Taras Shevchenko National University of Kyiv, Ukraine
          }

\abstract{
Given the increasing data collection capabilities and limited computing resources of future collider experiments, interest in using generative neural networks for the fast simulation of collider events is growing. In our previous study, the Bounded Information Bottleneck Autoencoder (BIB-AE) architecture for generating photon showers in a high-granularity calorimeter showed a high accuracy modeling of various global differential shower distributions.  In this work, we investigate how the BIB-AE encodes this physics information in its latent space. Our understanding of this encoding allows us to propose methods to optimize the generation performance further, for example, by altering latent space sampling or by suggesting specific changes to hyperparameters. In particular, we improve the modeling of the shower shape along the particle incident axis.
}
\maketitle
\section{Introduction}

High-quality simulations of fundamental processes and particle interactions with complex detectors are crucial to data analysis in high energy physics.
Especially in the context of increasing data volumes from upcoming runs of the Large Hadron Collider (LHC) and future experiments, the production of datasets using Monte-Carlo-based simulators is increasingly becoming a computing bottleneck~\cite{atlas_fastsim}.

A way to accelerate simulations is based on generative machine learning models that leverage recent advances in computer vision and are implemented parallelizable on graphic processing hardware. Such fast simulations based on Generative Adversarial Networks (GANs)~\cite{GAN} for calorimeter physics were first introduced in Ref.~\cite{CaloGAN2} and have seen active development in recent years~\cite{CaloGAN, CaloGAN3, ErdmannWGAN, ErdmannWGAN2, ATLAS_Gen, ATLAS_Gen2, ATLAS_Gen3, Sofia}.
This approach starts with a small dataset obtained using classical simulation techniques and aims to amplify its usable statistics by training a generative model. 
The principal feasibility of amplification was shown in Ref.~\cite{Butter:2020qhk}.

Inspired by Ref.~\cite{BIB-AE}, we have previously implemented an improved Bounded Information Bottleneck Autoencoder (BIB-AE) architecture and shown its generation accuracy for various differential distributions of photon shower data in a high granularity calorimeter~\cite{Getting_High}.
The BIB-AE architecture unifies ideas from different generative approaches, including GANs and Variational Autoencoders (VAE)~\cite{VAE}. 
As an autoencoder, the model encodes input photon showers into a latent space from which in turn newly generated showers are sampled.
The information bottleneck (IB) \cite{tishby2000information} refers in this context to the principle that the model optimizes the latent encoding while maximizing the mutual information between input and output.
This contribution explores methods to understand the physics encoded in the latent space and introduces optimizations for improved generation fidelity.
As opposed to Ref.~\cite{Howard:2021pos} we do not aim to explicitly shape the latent space to match physical distributions but rather investigate how the deviations and correlations of the optimally Gaussian normal latent space features correspond to physically important observables. Compared to Ref.~\cite{Batson:2021agz} 
we focus on an information-theoretic perspective and investigate correlations
with physical observables instead of the topological structure of the latent space. 

In the following, we first briefly introduce the data (Sec.~\ref{sec:dataset})
and our BIB-AE architecture (Sec.~\ref{sec:model}). We then investigate the connection between generative performance and information encoded in the latent space in Sec.~\ref{sec:diff_latent_sizes}, the correlation between learned latent space distributions and physical observables in Sec.~\ref{sec:corr_lat_phys}, and see how this can be used
to improve generative performance in Sec.~\ref{sec:improving_generative_performance}. We close with a summary of results and draw our conclusions in Sec.~\ref{sec:summary}.



\subsection{Photon showers in a high granularity calorimeter}
\label{sec:dataset}

Calorimeters are an essential part of detectors used at high-energy particle colliders.
They measure the energy particles deposit when interacting with material. Particles interacting with the matter in the calorimeter can produce secondary particles resulting in cascades or \textit{showers}. Such a particle shower is created for example by an initial electromagnetically interacting photon.

Modern sampling calorimeters are built in a sandwich structure of measuring active layers interspersed with dense passive material.
The active material of modern high granularity calorimeters consists of many small cells that are read out separately, yielding high resolution 3-dimensional measurements of particle showers.

We created our photon shower dataset using the \textsc{Geant4}~\cite{g4} toolkit and a simulation of the SiW electromagnetic calorimeter in the International Large Detector (ILD) concept~\cite{ILD-IDR}. The simulated calorimeter section comprises 30 active layers with each 900 5x5\,$\textrm{mm}^2$ calorimeter cells in a rectangular grid resulting in 3d images of $30 \times 30 \times 30 = 27,000$ pixels.
Our dataset consists of 950k photon showers with incident energies uniformly distributed between 10 and 100\,GeV. This is the same dataset as used for Ref.~\cite{Getting_High} and we refer to that publication for additional details.~\footnote{A fraction of the dataset as well as our implementation of the BIB-AE model are available at \\
\url{https://github.com/FLC-QU-hep/getting_high}.}

\subsection{The BIB-AE model}
\label{sec:model}

\begin{figure*}[t]
    \centering
    \includegraphics[width=0.9\textwidth]{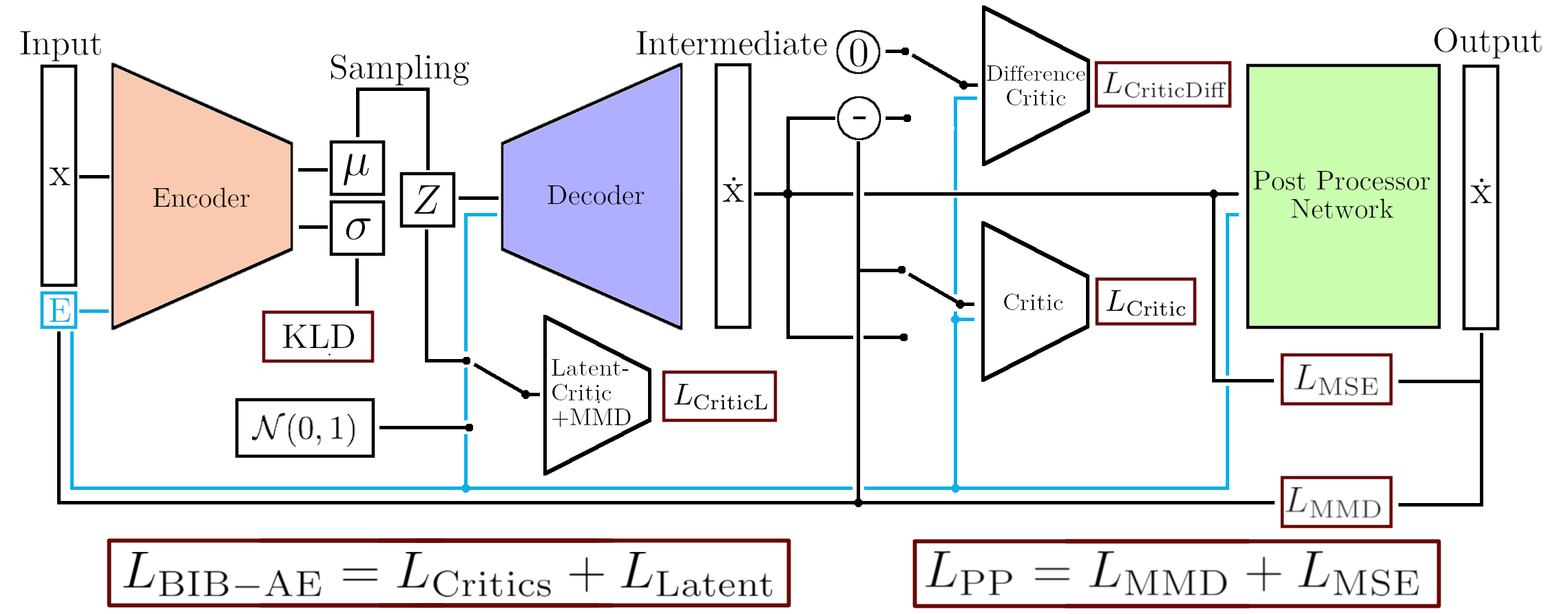}
    \caption{Overview of the BIB-AE generative model including the Post-Processor (PP) network and their respective loss terms.
    The model consists of multiple networks: An encoder, a decoder, a Post-Processor network as well as two reconstruction critics and a latent space critic. The latent critic, the Kullback-Leibner divergence (KLD) and a latent MMD regularize the latent space towards a Standard Normal distribution.
    The BIB-AE PP is conditioned on the incident particles' energy (blue lines).
    }
    \label{fig:architecture_BIBAE_PP}
\end{figure*}

The BIB-AE architecture consists of several building blocks: An encoder network mapping the input calorimeter images into a latent representation; a decoder network transforming the latent representation back into calorimeter images; a Post-Processor network refining the pixel values of the decoded image; a reconstruction critic network calculating the Wasserstein-distance between encoded and decoded image; and a latent critic network to regularize the latent space. 
The whole model is trained in two stages: First, the encoder, decoder, and critics are trained until sufficient fidelity is reached; afterwards, the whole model is trained in conjunction with the Post-Processor network to improve the accuracy of the cell energy generation.
To generate energy dependent samples, the BIB-AE is conditioned on the incident particle energy.
An overview of the architecture is shown in Figure~\ref{fig:architecture_BIBAE_PP}. A more detailed discussion of the network is provided in \cite{Getting_High}.


Like in any VAE-based model, the trained BIB-AE can be used to generate calorimeter shower images by sampling the latent space from Standard Normal distributions. To  achieve good generation results, the latent space needs to be regularized towards such a Normal distribution. For this regularization the BIB-AE model employs several loss terms during training: A Kullback-Leibler divergence (KLD) loss $L_{\textrm{KLD}}$, the output of a latent critic network $L_{\textrm{latent-critic}}$, and a latent Maximum Mean Discrepancy (MMD)~\cite{MMD_base} term $L_{\textrm{latent-MMD}}$. Each latent regularizer contribution is scaled with a weight $\beta$ yielding a combined latent loss of
\begin{equation}
    \label{eq:BAE_latent_loss_total}  
    L_{\textrm{total-latent}} = \beta_{\textrm{KLD}} \cdot  L_{\textrm{KLD}} + \beta_{\textrm{latent-critic}} \cdot L_{\textrm{latent-critic}} + \beta_{\textrm{latent-MMD}} \cdot L_{\textrm{latent-MMD}}
\end{equation}
with the KL divergence of two discrete probability distributions $P$ and $Q$ defined as 
\begin{equation}
    \label{eq:KLD_definition}
    D_\text{KL}(P \parallel Q) = \sum_{x\in\mathcal{X}} P(x) \log\left(\frac{P(x)}{Q(x)}\right)
\end{equation}
and calculated via
\begin{equation}
    \label{eq:KLD_calculation}
    D_{\textrm{KL}, i} = D_{\textrm{KL}}(\mathcal{Z}_i \parallel \mathcal{N}(0,1)) = - \frac{1}{2} \left( 1 + \text{log}(\sigma_i^2) - \mu_i^2 - \sigma_i^2 \right).
\end{equation}

In the context of this publication, \textit{latent variables} $\mathcal{Z}_i$ are Gaussian distributions with two trainable parameters $\mu_i$ and $\sigma_i$ ($\mathcal{Z}_i \equiv \mathcal{N}(\mu_i,\sigma_i^2)$)
regularized towards a Standard Normal distribution.
Its sampled values are $z_i \sim \mathcal{Z}_i$. 


In previous work we have implemented the BIB-AE architecture with 24 trainable latent variables and an additional 488 variables that are not encoded but sampled straight from a Standard Normal distribution during training~\cite{Getting_High}.
We term the number of trainable latent space variables the \textit{latent space size} $n$. Hence the total KLD loss is given by \mbox{$L_{\textrm{KLD}} = \sum_{i=1}^n D_{\textrm{KL}, i}$}.

The loss weight $\beta_{\textrm{KLD}}$ has the highest impact on the latent regularization as its scaling defines the magnitude of the KL divergence. Here the KL divergence measures the information content
of the latent space~\cite{shannon_information, info_theory_and_statistics}.


\section{Different latent space sizes}
\label{sec:diff_latent_sizes}

Intuitively, for fixed $\beta_{KLD}$, higher latent space sizes $n$ should yield an increased total information in the latent space until a maximum corresponding to the showers' intrinsic relevant information is reached.
We test this by re-training a BIB-AE architecture with latent space sizes between 2 and 512 for fixed $\beta_{KLD}=0.05$.
The number of additional Standard Normal sampled variables is adjusted such that the total number of 512 decoded variables stays constant.

\begin{figure*}[t]
    \centering
    \includegraphics[width=0.45\textwidth]{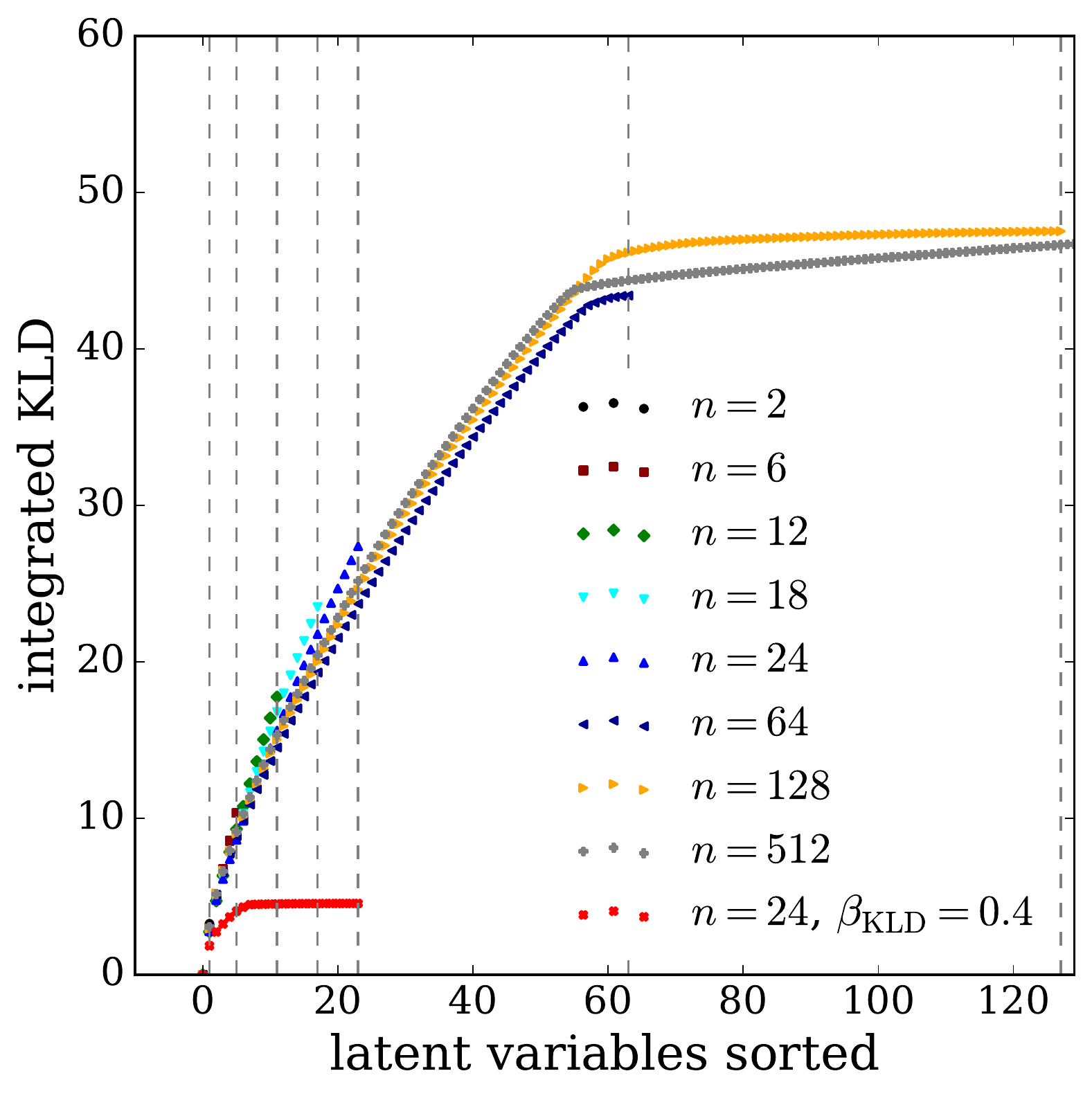}
    \includegraphics[width=0.45\textwidth]{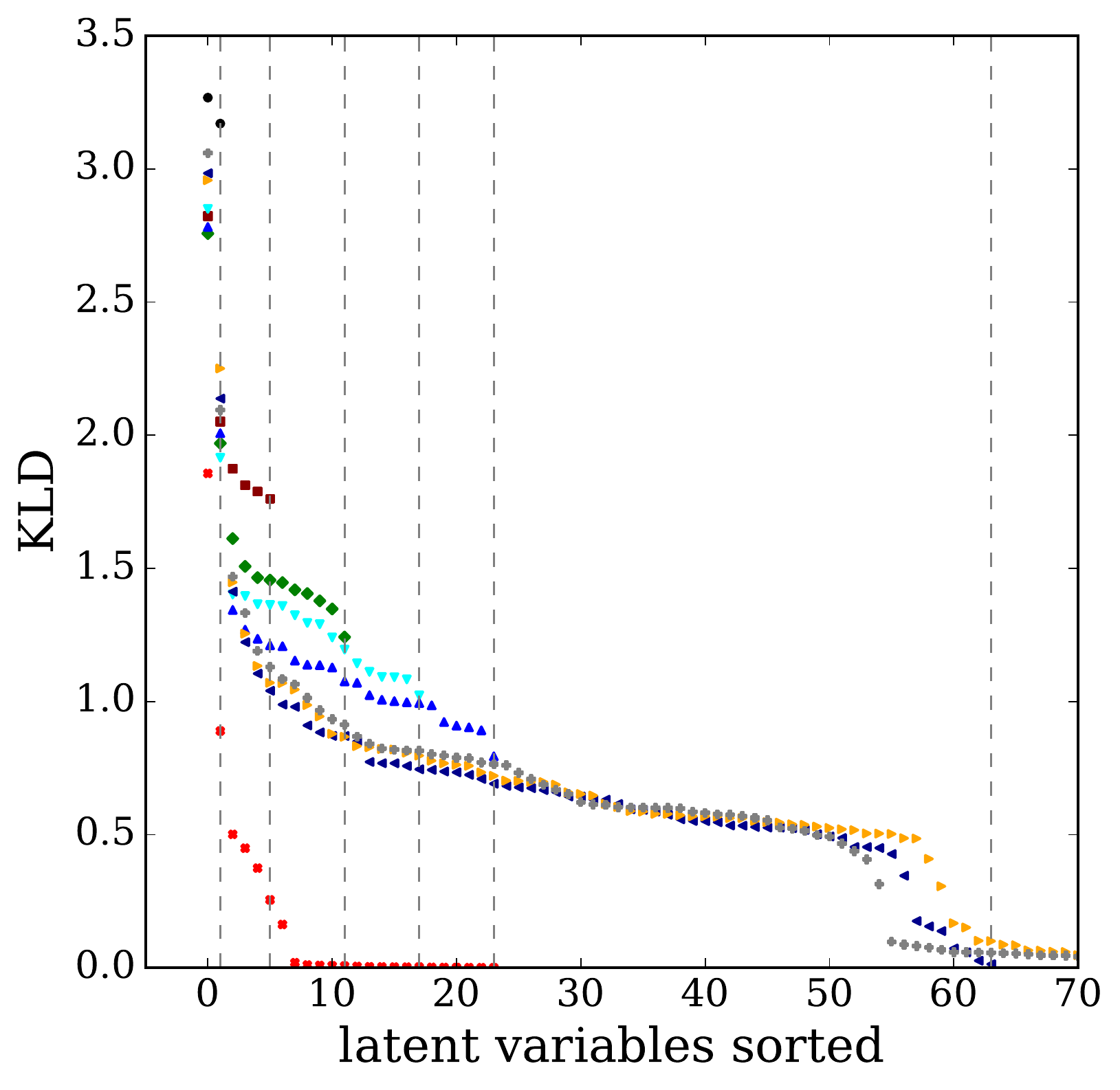}
    \caption{
    \textbf{Left:} Integrated Kullback-Leibler divergence (KLD) for latent variables sorted by highest KLD for models with different latent space sizes.
    \textbf{Right:} KL divergence of individual latent space variables sorted with decreasing KLD for different latent space sizes. All models are trained with a baseline weight $\beta_{\textrm{KLD}}=0.05$, except stated otherwise.
    }
    \label{fig:KLDvsLatVarSort}
\end{figure*}

In Fig.~\ref{fig:KLDvsLatVarSort} (left) we have sorted the trainable latent space variables by their individual KLD calculated via Eq.~\ref{eq:KLD_calculation}. 
On the vertical axis, the total information (i.e. the sum of KLD values up to and including latent variable $i$) is shown. Indeed, we observe increasing total encoded information with increasing latent space size until a saturation at approximately 45~nats ($\approx 64$~bits)
is reached around a latent space size of 64. After that, a larger latent space does not substantially increase the learned information.

Next, we consider the KLD per latent space variable in  Fig.~\ref{fig:KLDvsLatVarSort}~(right).
All models follow a similar pattern: Only a few variables contain a high amount of information (high KLD). 
In particular, there are always two variables that encode significantly more information than the remaining ones. Furthermore, about 60 variables contain $> 0.3$~nats of information depending on the latent space size. 


Naively, we would expect the most efficient use of the latent space at a size of $n=64$ to yield optimal performance. Evaluating the performance of generative models is not straightforward and constitutes an active topic of research. Methods such as \textit{Inception Score}~\cite{inception_score} were proposed to evaluate models which produce photographic images.  
However, such scores are typically domain-specific and cannot directly be applied to our dataset.  
We therefore define a problem specific \textit{fidelity score} $S_{\textrm{JSD}}$ that summarizes the performance across a number of physically relevant observables. The score is calculated by combining the Jensen–Shannon distance (JSD) between the \textsc{Geant4} truth and generation results of the six one dimensional histograms shown in Fig.~\ref{fig:plots_KLD005vs04}, namely the visible cell energy, the total shower energy, the occupancy, the center of gravity in z as well as the radial and longitudinal energy distributions. 
These are some of the most relevant global differential distributions for photon shower analysis and were applied previously to judge model performances~\cite{Getting_High}.
Additional details on how the score is calculated are given in  Appendix~\ref{app:fidelity_metric}. 

In Table~\ref{tab:fidelity_score_bestEp} we show the fidelity score for different values of the latent space size $n$. 
Lower values correspond to better agreement with the underlying slow simulation. For very small $n$, the performance
increases with $n$ until the best observed value at $n=24$. 
Seven trainings with identical network setup but different random weight initialization were performed for this point to obtain an estimate of the associated uncertainty (calculated as the standard deviation of individual results at $n=24$). 
Further hyperparameters were kept the same as in Ref. \cite{Getting_High}.
In the table the best score out of those trainings is given.
Limited computing resources due to several days of training needed per model did not allow for a wider estimation of the fidelity scores. 
For larger $n$ the performance is approximately stable within the
uncertainty observed for $n=24$.
This implies that maximum information content of $\approx 45$~nats encoded in the latent space is not needed for optimal generative performance.


\begin{table}
\centering
\caption{Fidelity score $S_{\text{JSD}}$ for the best epoch of multiple model configurations with different latent space sizes $n$. 
For $n=24$ the best score out of multiple training runs is given, while the mean score for those trainings is: $\overline{S}_{\text{JSD,24}} = 1.02 \pm 0.12$.
Only one training each  was performed for sizes $n \neq 24$.
}
\label{tab:fidelity_score_bestEp} 
\begin{tabular}{lcccccccc}
\hline
\textbf{latent size} & 2 & 6 & 12 & 18 & 24 & 64 & 128 & 512
\\ \hline
\textbf{$S_{\text{JSD}}$} & 1.64 & 1.12 & 1.11 & 0.95 & \textbf{0.83}  & 0.88 & 0.94 & 0.98  \\ \hline
\end{tabular}
\end{table}


\section{Correlations between latent space and physics}
\label{sec:corr_lat_phys}

\begin{figure*}[t]
    \centering
    \includegraphics[width=0.2776\textwidth]{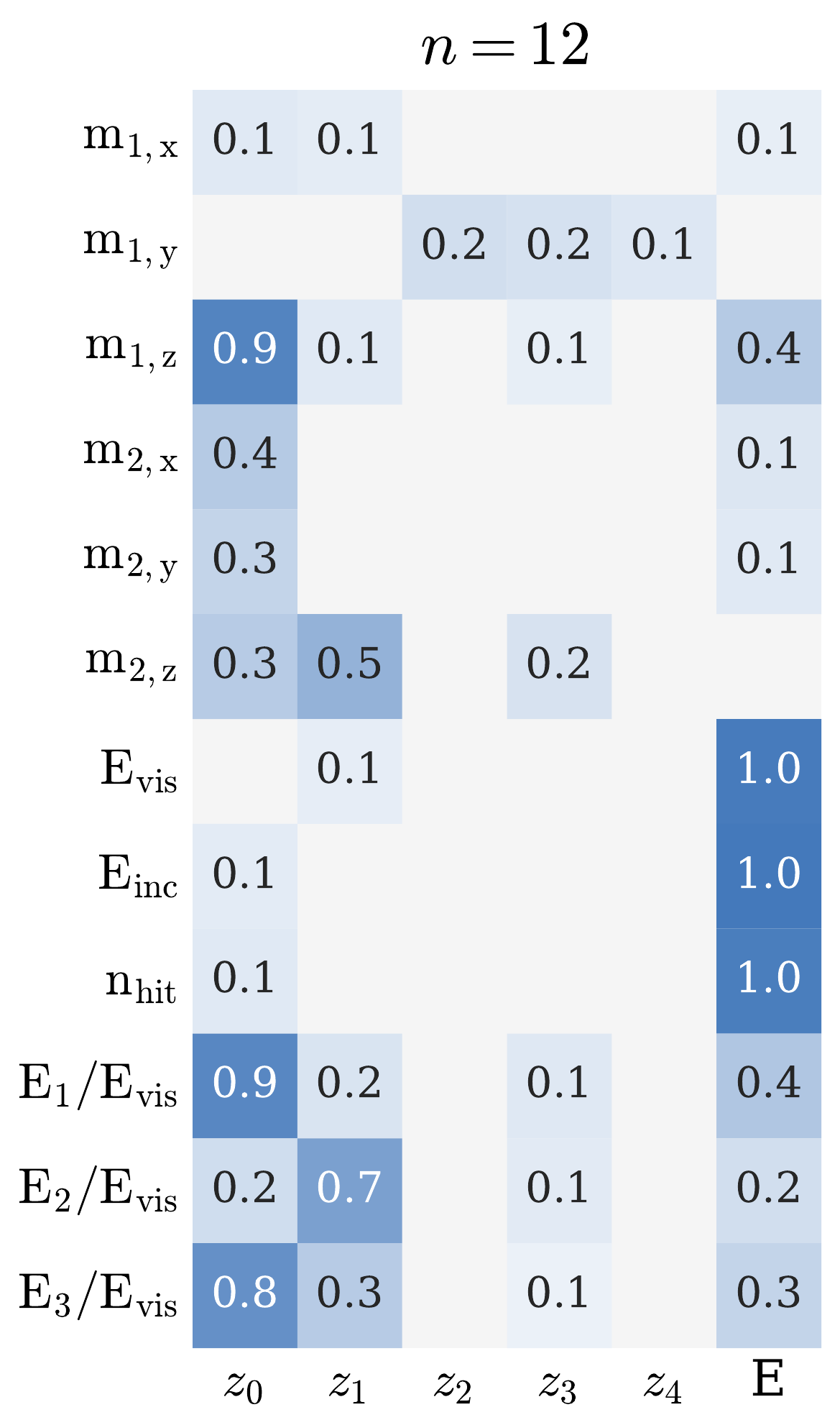}
    \includegraphics[width=0.22\textwidth]{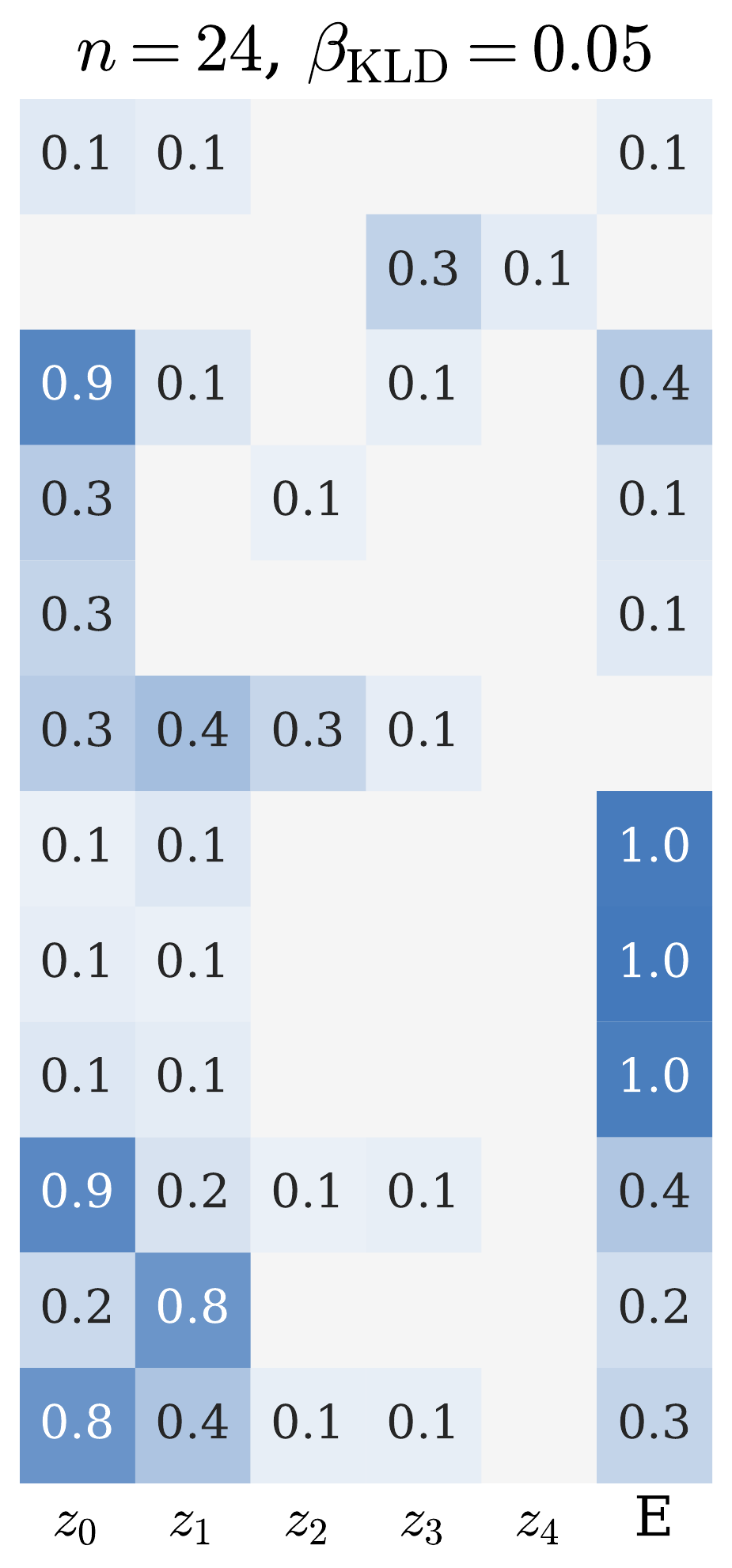}
    \includegraphics[width=0.22\textwidth]{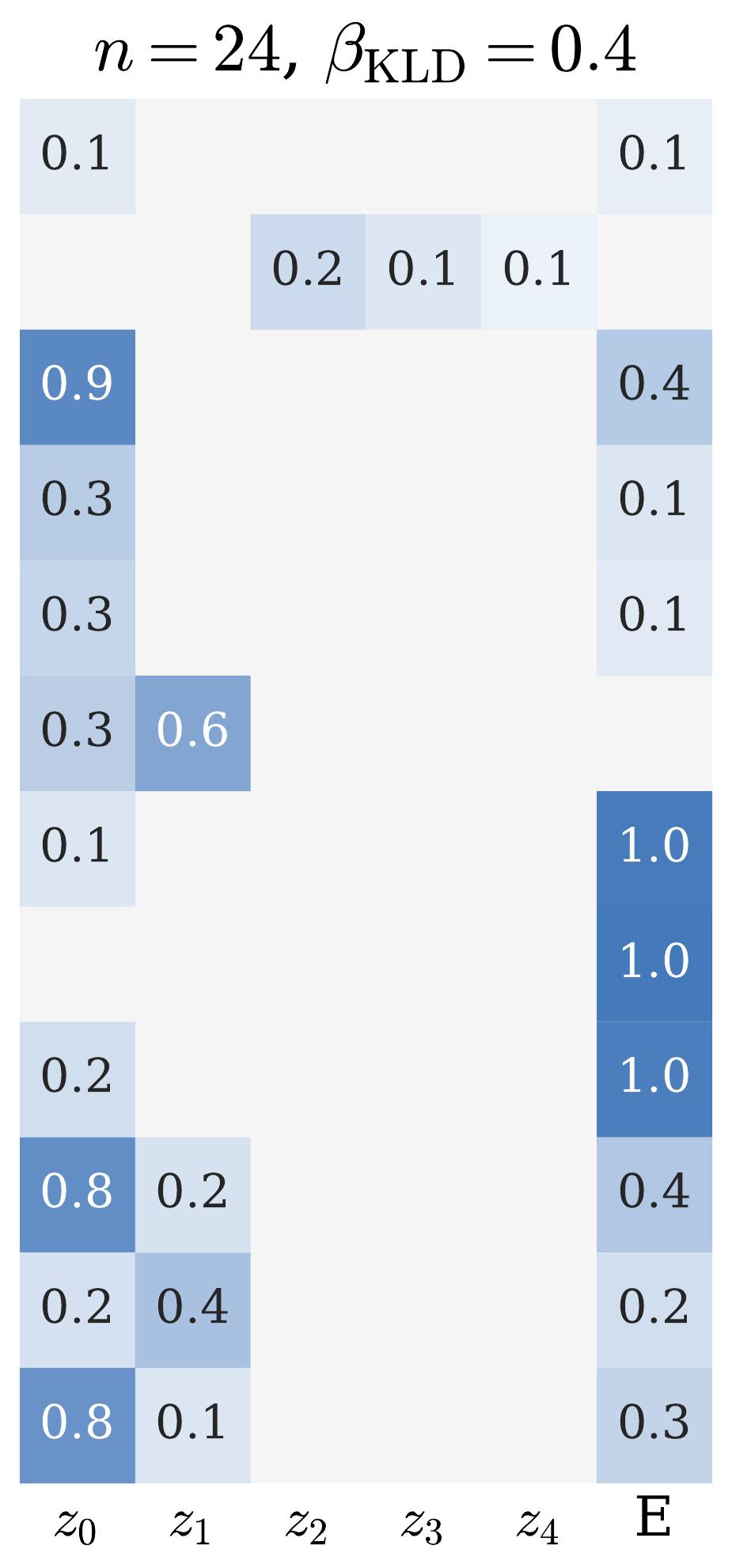}
    \includegraphics[width=0.22\textwidth]{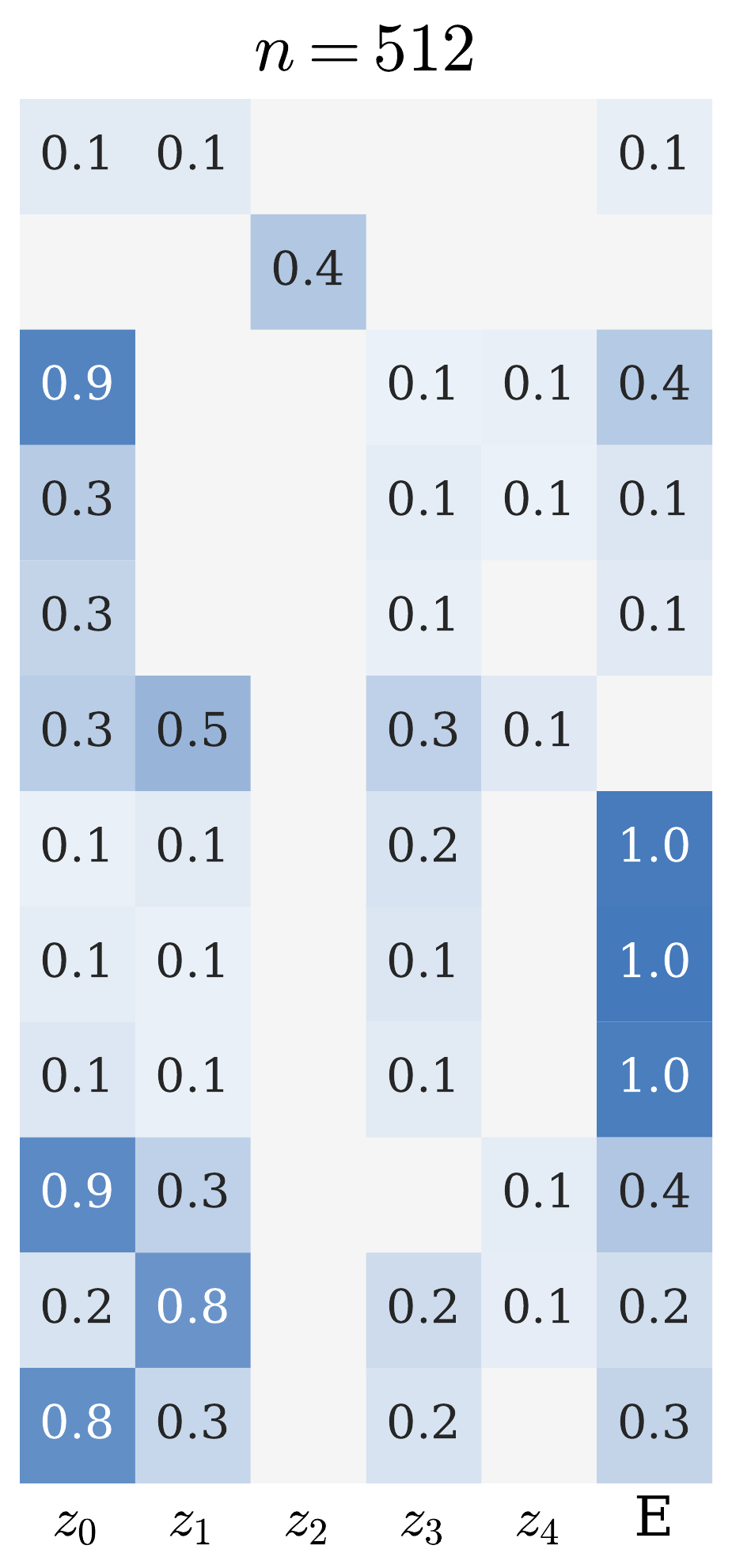}
    \caption{Pearson correlation coefficients between various physics variables and the sampled $z_i$ of the five highest KLD latent variables as well as the incident particle energy E for multiple model sizes $n$. The baseline latent weight is $\beta_{KLD} = 0.05$ except for one training with $\beta_{KLD} = 0.4$.
    Only non-zero values of the  correlations are shown.
    }
    \label{fig:correlations_KLD005vsKLD04}
\end{figure*}

As only a few variables seem to encode most of the shower information, we investigate what kind of physics information is learned by these variables.
In Fig.~\ref{fig:correlations_KLD005vsKLD04} the Pearson correlation coefficients $\rho$ between different shower physics distributions and the distributions of the sampled $z_i$ for the five highest KLD latent variables as well as the incident particle energy (which is used for conditioning and is included as a latent variable in the BIB-AE) are shown for four different model configurations. In this case the sampled $z_i$ values are obtained from the encoded latent space via $\mathcal{N}(\mu_i, \sigma_i^2)$, not from a Standard Normal distribution $\mathcal{N}(0,1)$.
The physics distributions include the first and second moment in each spatial dimension~\footnote{The incident photon enters the calorimeter in the center of the x-y plane at z=0 and traverses along the z-axis.} --- the first moment corresponds to the center of gravity --- the visible energy, the incident particle energy, the number of hits, and the fraction of deposited energy in each third of the calorimeter (in the z-direction). 

Regardless of the model configuration, it is apparent that the highest KLD latent variable strongly correlates (approx. $\rho$ = 0.9) with the center of gravity along the shower direction z (and in turn to the fraction of deposited energy in the first and last third of the calorimeter). Another variable is correlated (approx. $\rho$ = 0.5) to the second moment in z (and to the energy fraction in the middle of the calorimeter). It appears that of all the shower variables, the center of gravity in z (CoG-Z) of each shower is encoded into these two latent variables. This is important as the CoG-Z is not as much correlated to the incident particle energy (approx. $\rho$ = 0.4) on which the BIB-AE is conditioned. Hence the BIB-AE learns the CoG-Z of each shower and uses it in the decoding step for reconstruction. Interestingly, this pattern is very stable over multiple independent training runs and even different latent space sizes $n$.

\begin{figure*}[t]
    \centering
    \includegraphics[width=0.22\textwidth]{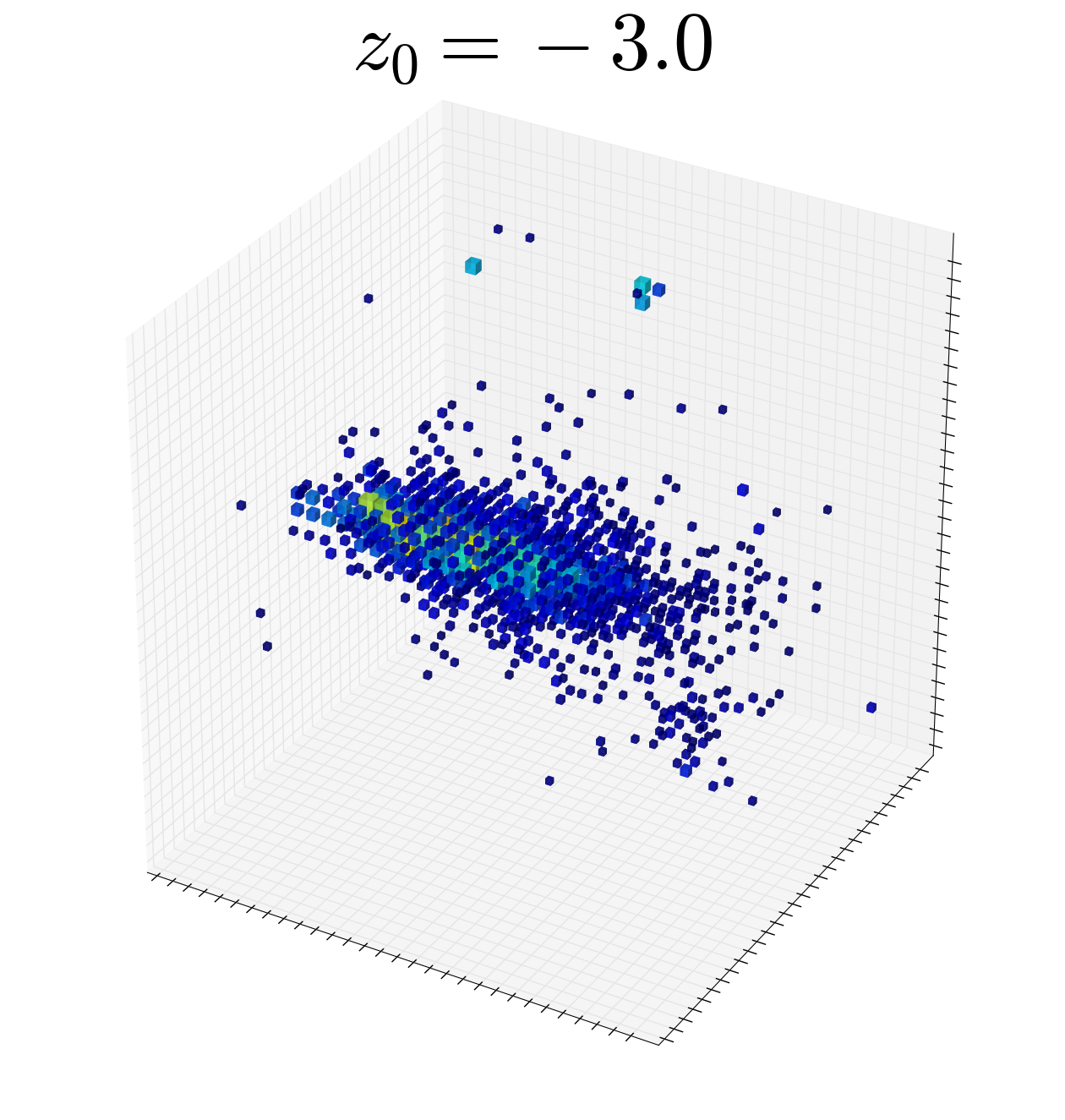}\hspace{-0.5cm}
    \includegraphics[width=0.22\textwidth]{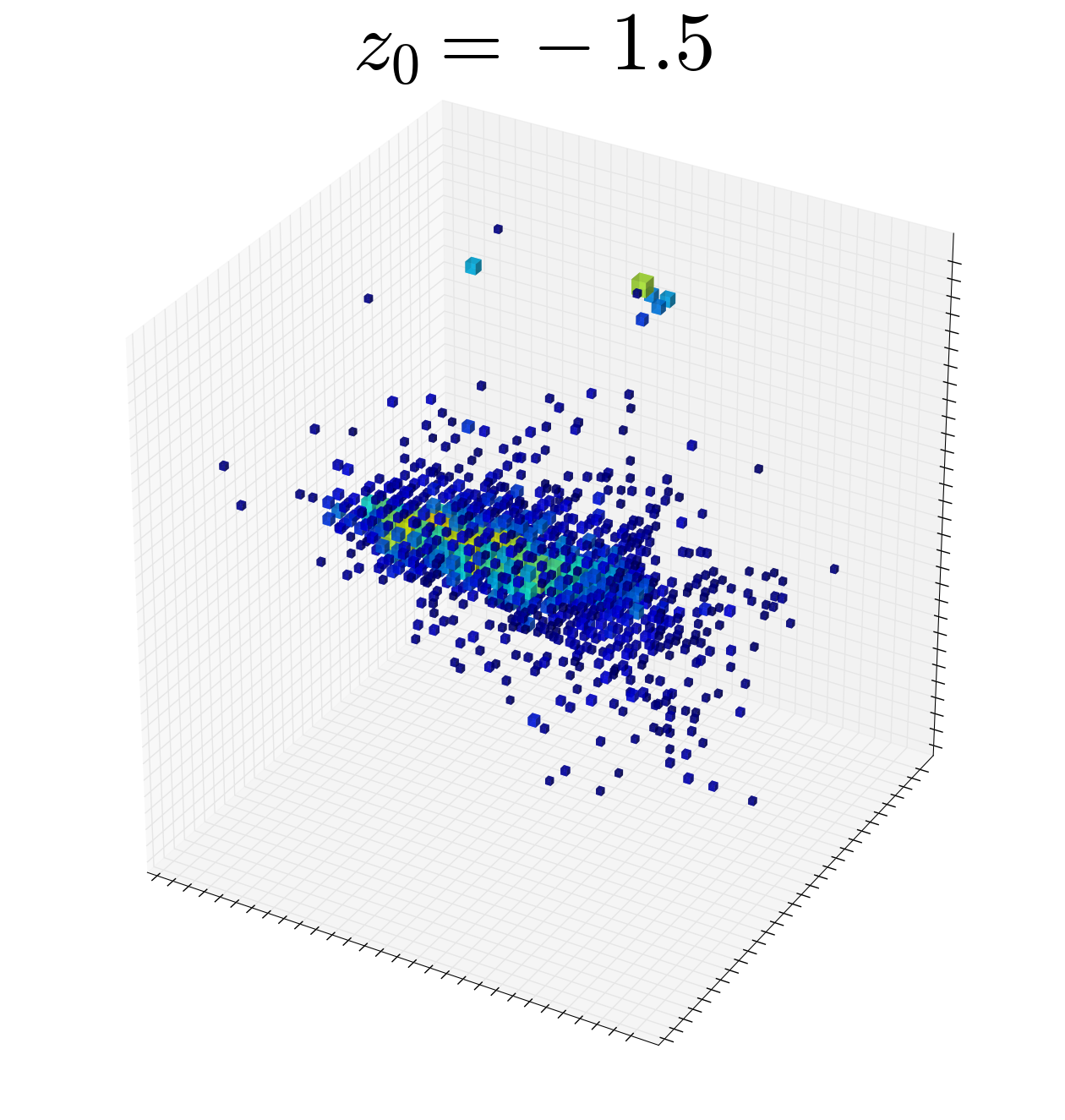}\hspace{-0.5cm}
    \includegraphics[width=0.22\textwidth]{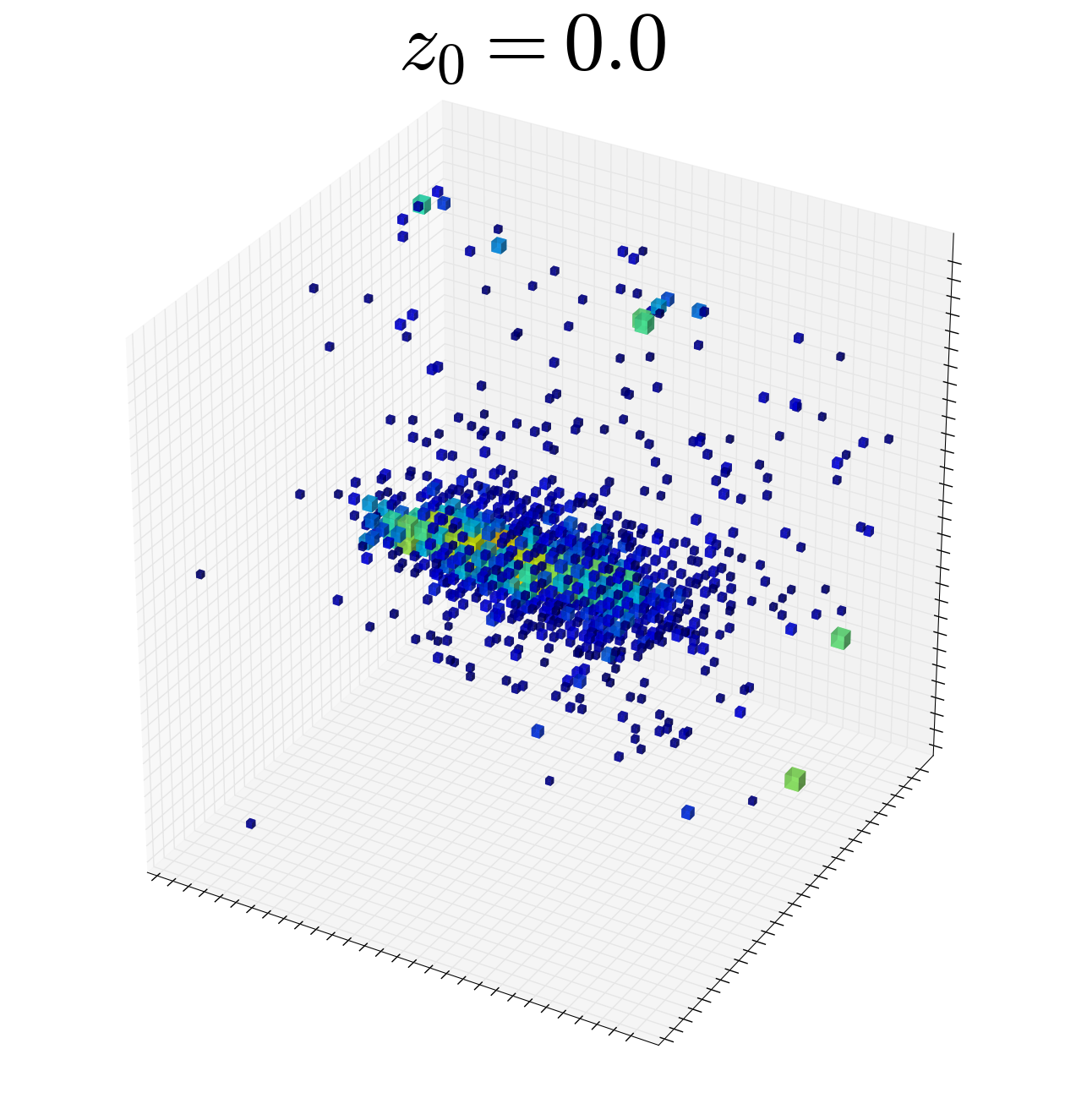}\hspace{-0.5cm}
    \includegraphics[width=0.22\textwidth]{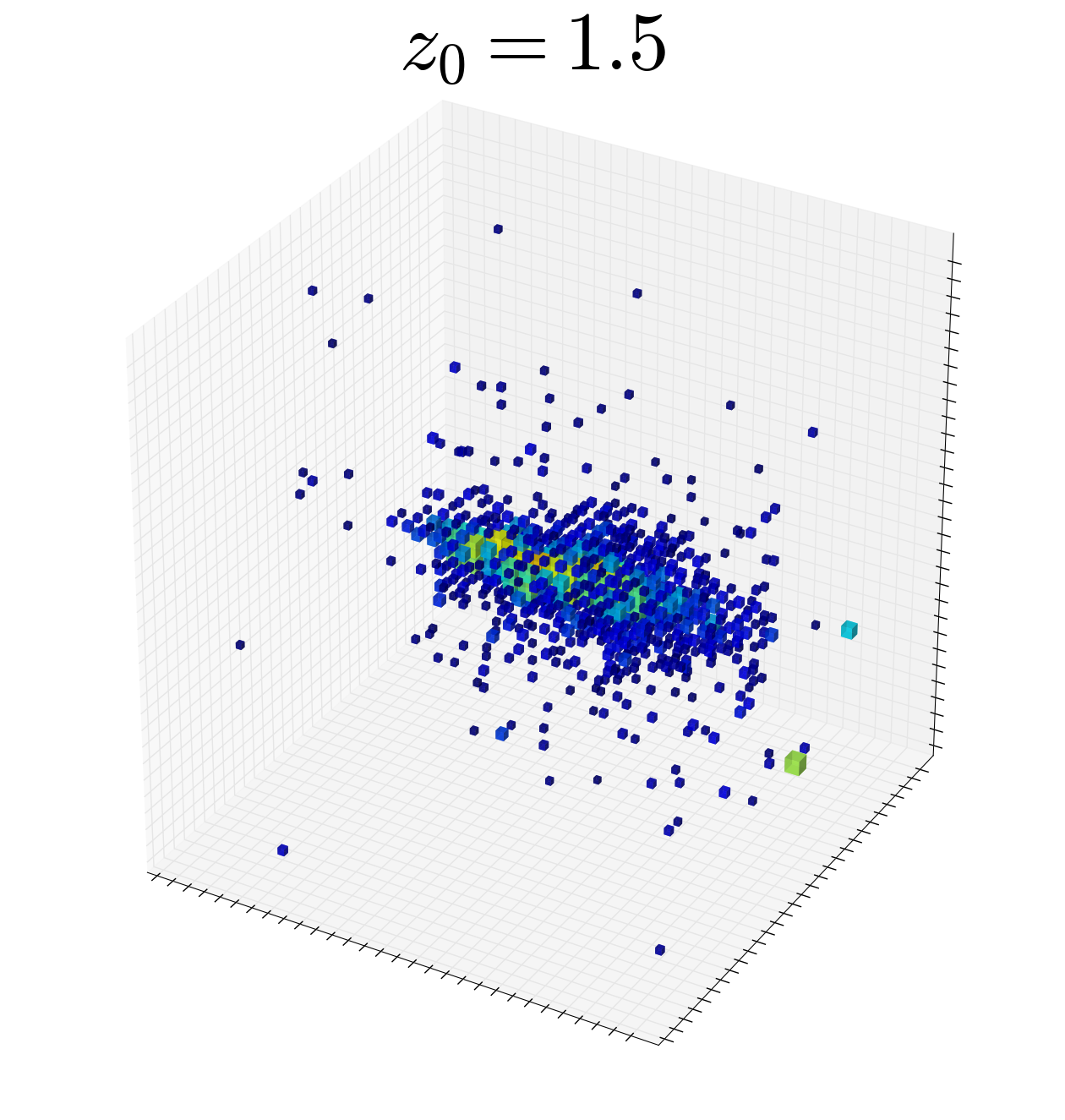}\hspace{-0.5cm}
    \includegraphics[width=0.22\textwidth]{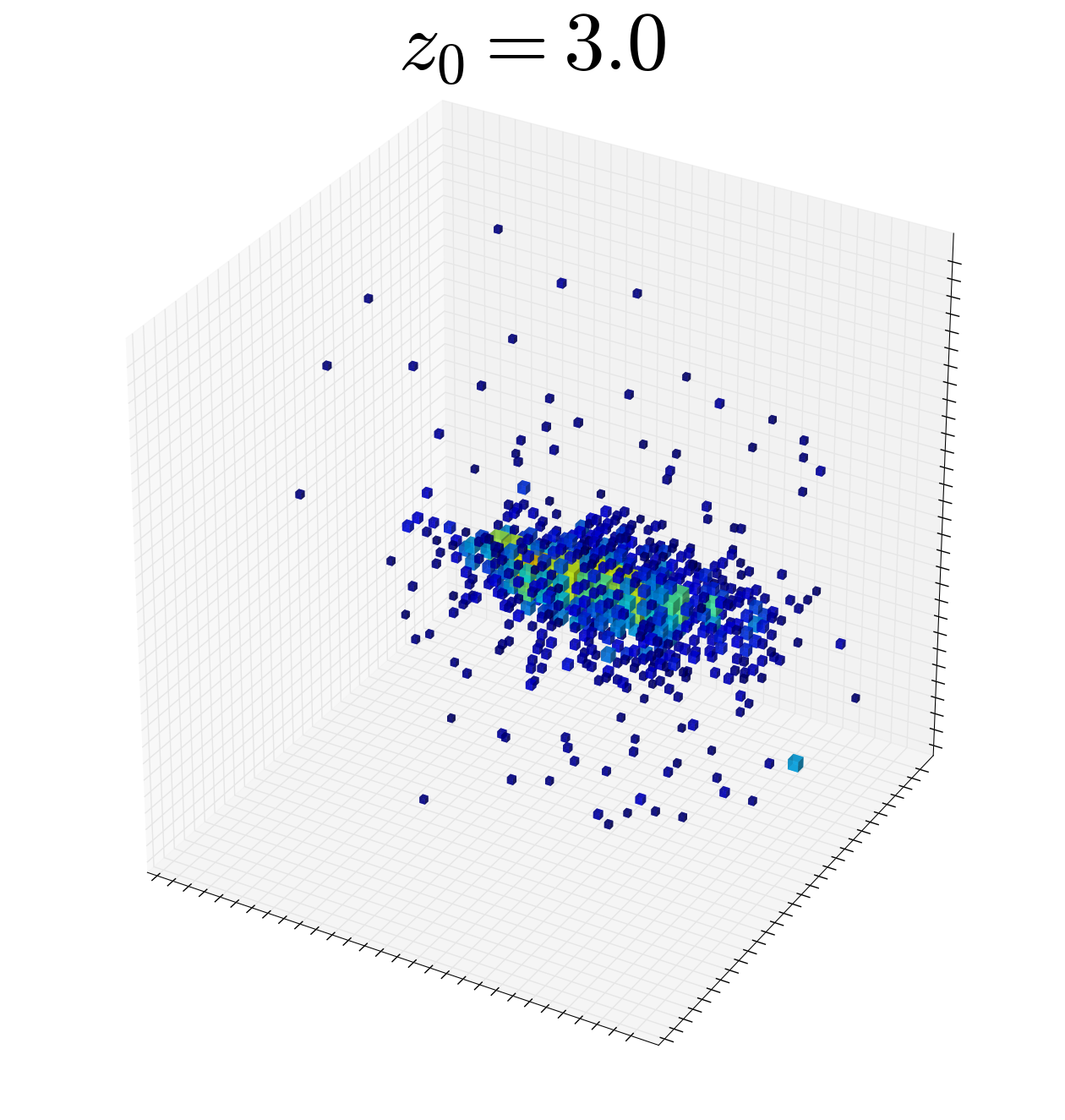}
    \caption{3d image of generated showers decoded from a latent space with all variables $z_i= 0 $, except the highest KLD latent $z_0$ variable which is set to values between -3 and 3.
    The color coding corresponds to the cells' energy deposition. The highest KLD variable $z_0$ correlates to the CoG-Z distribution, hence an evolution of the shower start is visible (projection found in Fig.~\ref{fig:layerZ_augLat0}).
    }
    \label{fig:latent_scan_3d}
\end{figure*}

\begin{figure*}[t]
    \centering
    \sidecaption
    \includegraphics[width=0.7\textwidth]{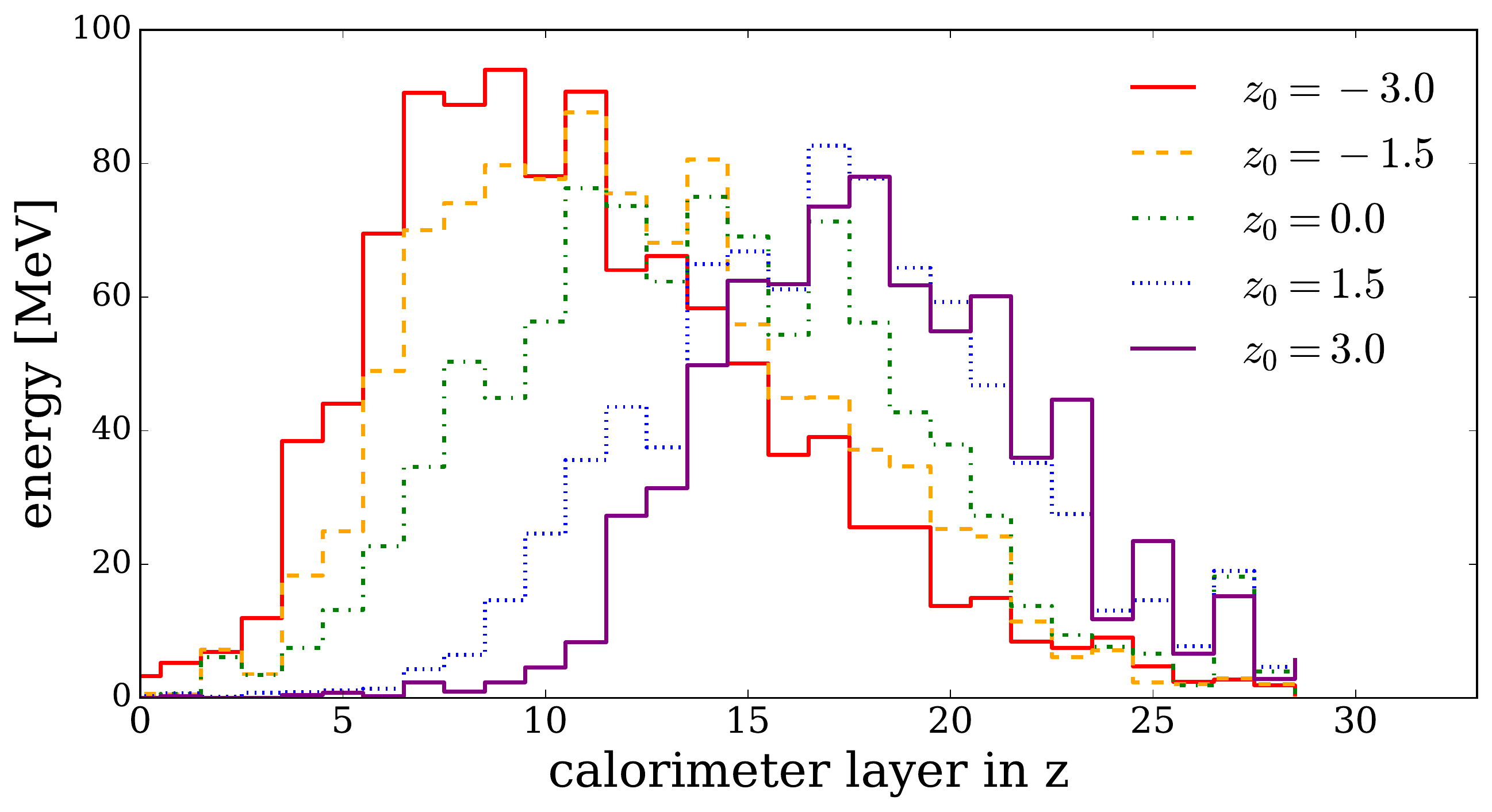}
    \caption{Deposited energy per layer in z-direction for showers which are decoded with all latent variables $z_i = 0$, except the highest KLD latent $z_0$ variable which is set to values between -3 and 3.  
    }
    \label{fig:layerZ_augLat0}
\end{figure*}

\begin{figure*}[ht]
    \centering
    \includegraphics[width=0.31\textwidth]{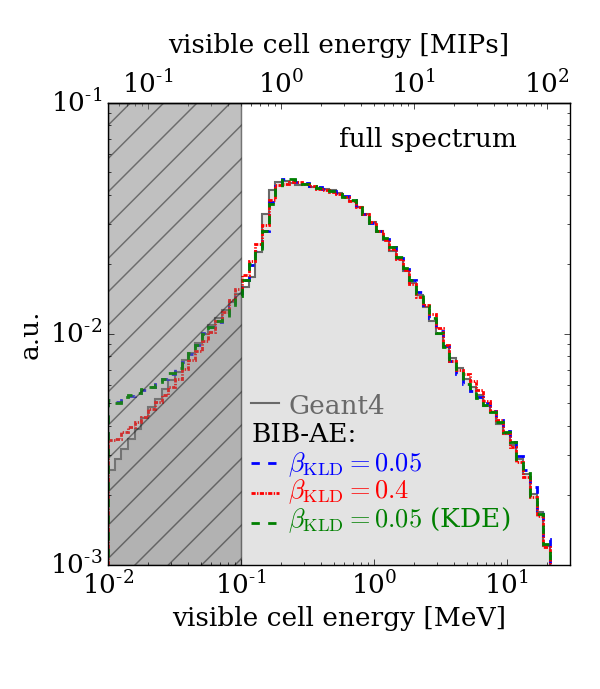}
    \includegraphics[width=0.31\textwidth]{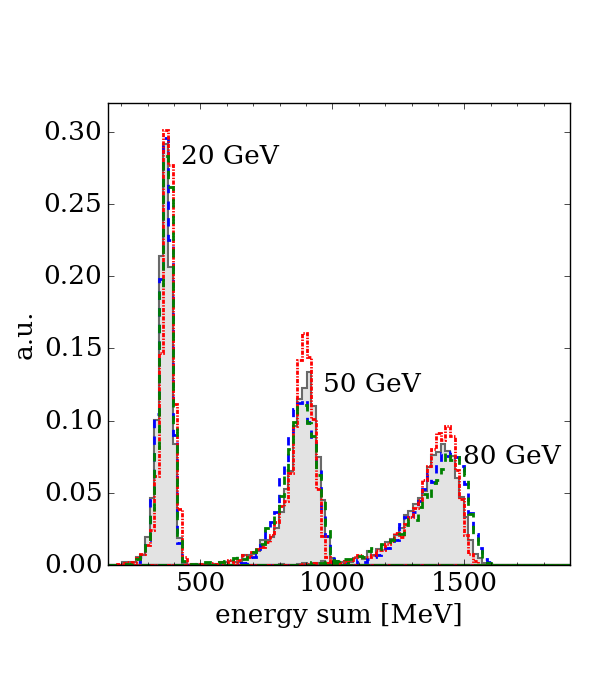}
    \includegraphics[width=0.31\textwidth]{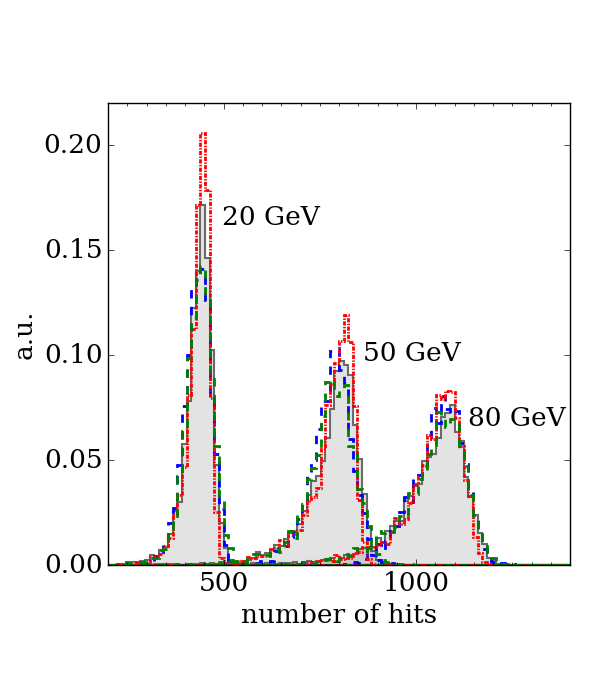}
    \includegraphics[width=0.31\textwidth]{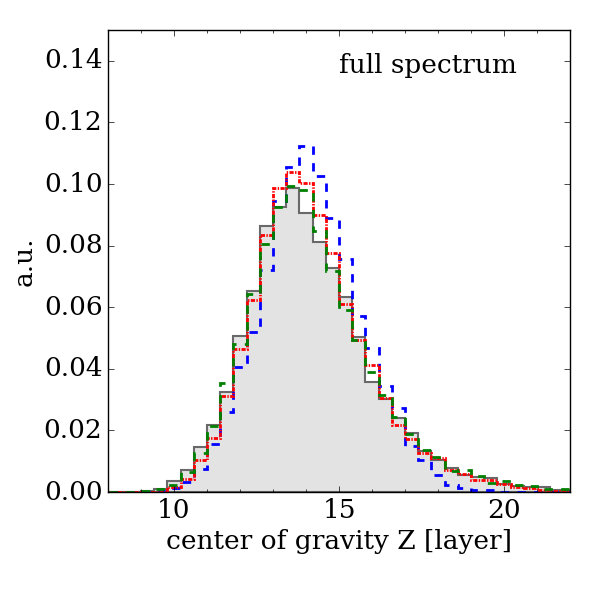}
    \includegraphics[width=0.31\textwidth]{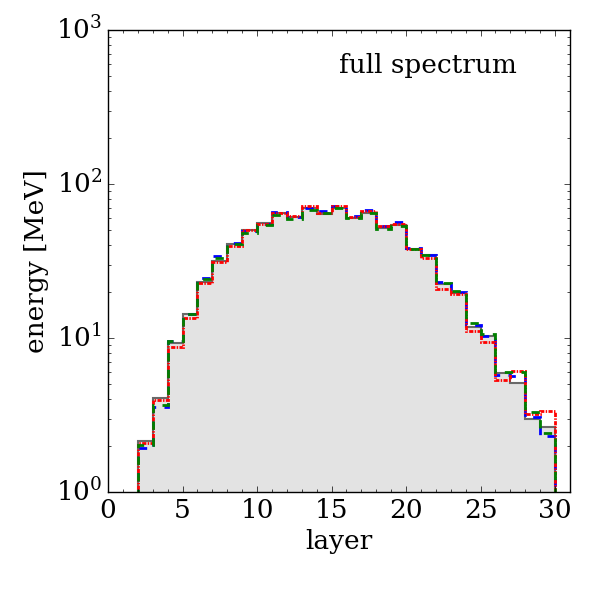}
    \includegraphics[width=0.31\textwidth]{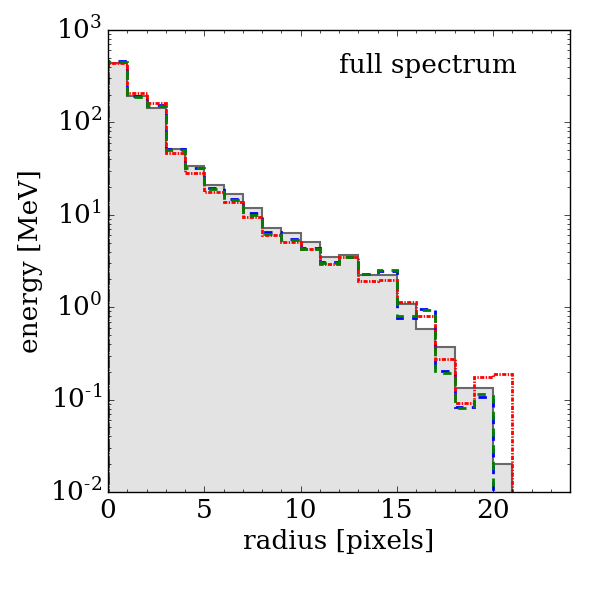}
    \caption{Differential distributions comparing physics quantities between \textsc{Geant4} and BIB-AE models with $\beta_{KLD} = 0.05$, $\beta_{KLD} = 0.4$ and $\beta_{KLD} = 0.05$ with the KDE sampling approach.
    }
    \label{fig:plots_KLD005vs04}
\end{figure*}

We can use this observation to improve the CoG-Z distribution in the generated events~(see Fig.~\ref{fig:plots_KLD005vs04} (bottom left)). This distribution was previously not particularly well-modeled since the generation did not take this latent space correlation into account.
In addition, the targeted sampling of a subspace of these latent variables allows to generate showers with a specific shower start.
This is visualized in Fig.~\ref{fig:latent_scan_3d} with multiple 3d images of a decoded calorimeter shower in which only the highest KLD latent $z$ variable was altered. This variable change leads to a different shower start and hence to an altered center of gravity in the z-axis. 
Figure \ref{fig:layerZ_augLat0} visualizes the energy deposition per layer in z-direction of these five decoded shower images. 

\section{Improving generative performance}
\label{sec:improving_generative_performance}

Understanding the encoded shower information, particularly the center of gravity, in the latent space helps us make educated optimization choices for improving model performance. 
Specifically, we can increase generation fidelity by either regularizing the latent space more strongly or by leveraging and sampling from the information rich non-Gaussian distributions. 
Either optimization path can be approached in different ways. 
We have chosen one exemplary method for each:
(1) By increasing $\beta_{KLD}$ the overall KLD in the latent space is reduced, yielding latent distributions stronger regularized towards Standard Normal distributions and therefore more accurate generative sampling from such a $\mathcal{N}(0,1)$ distribution; or (2) keeping the already trained model but using a second density estimator 
--- such as Kernel Density Estimation (KDE)~\cite{KDE} ---
on the latent variables and sampling directly from the encoded latent space.  
The former approach is motivated by \cite{beta-VAE} while the latter mirrors a method for the \textit{Buffer-VAE} from Ref.~\cite{Buffer_VAE}.

\begin{figure*}[ht]
    \centering
    \sidecaption
    \includegraphics[width=0.35\textwidth]{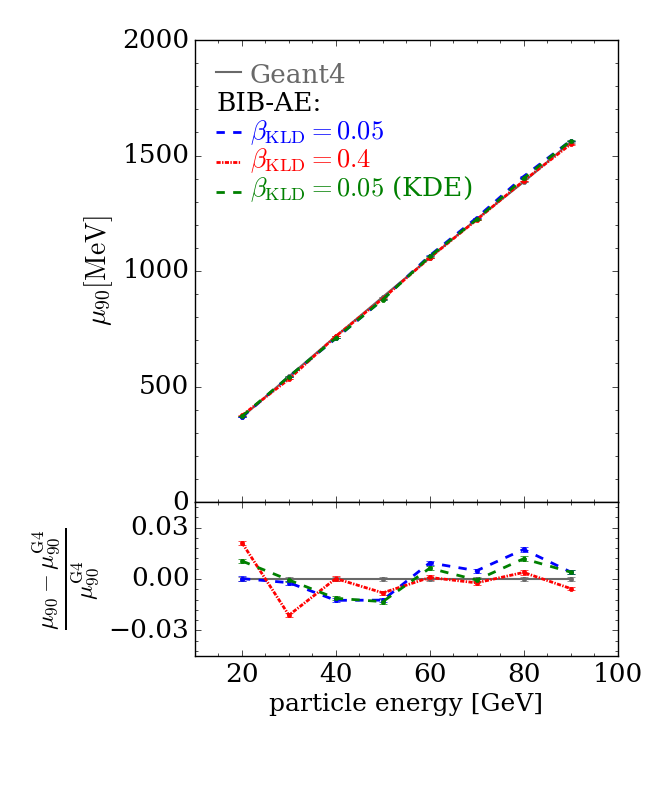}
    \includegraphics[width=0.35\textwidth]{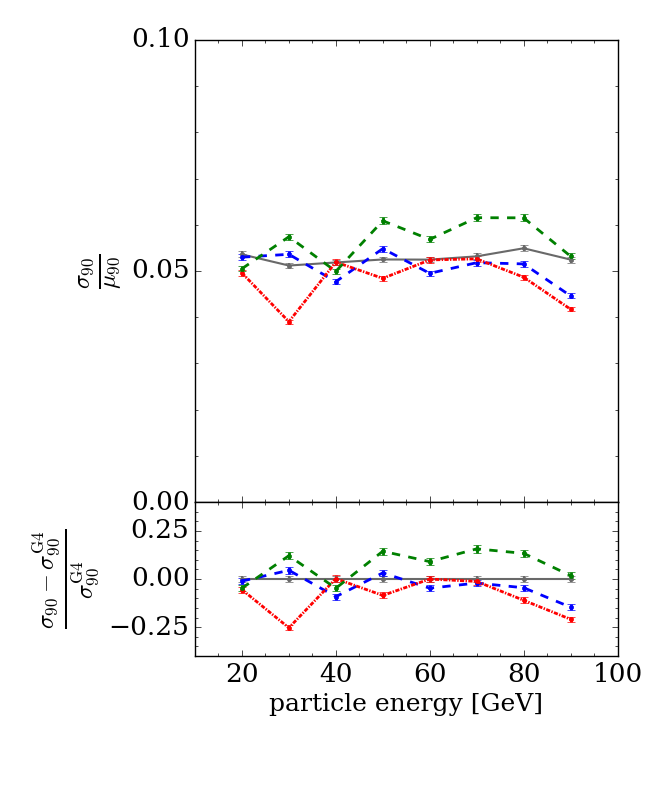}
    \caption{Mean and relative width of the energy deposited in the calorimeter for various incident particle energies for  \textsc{Geant4} and BIB-AE models with $\beta_{KLD} = 0.05$, $\beta_{KLD} = 0.4$ and $\beta_{KLD} = 0.05$ with the KDE sampling approach.
    }
    \label{fig:res_mean_KLD005vs04}
\end{figure*}

\begin{table}
\centering
\caption{Fidelity score $S_{\text{JSD}}$ for the best epochs for multiple model and sampling configurations of BIB-AE models with a latent size of 24.
For $\beta_{\textrm{KLD}}=0.05$ the best score out of multiple training runs is given, while the mean score for those trainings is: $\overline{S}_{\text{JSD,24}} = 1.02 \pm 0.12$.
For $\beta_{\textrm{KLD}}=0.4$ only one training was performed.
}
\label{tab:fidelity_scores_24} 
\begin{tabular}{lccc}
\hline
\textbf{config.} & $\beta_{KLD}=0.05$ & $\beta_{KLD}=0.4$ & $\beta_{KLD}=0.05$+KDE sampling
\\ \hline
\textbf{$S_{\text{JSD}}$} & 0.83 & 0.88 & \textbf{0.67}  \\ \hline
\end{tabular}
\end{table}

\subsection{Adjusting the Kullback-Leibler divergence}

Our baseline model uses a latent KLD weight of $\beta_{KLD} = 0.05$. However, as a higher value for $\beta_{KLD}$ leads to a lower KLD value, less information is encoded in the latent space. 
Therefore, the latent space more closely approaches a Standard Normal distribution and sampling from $\mathcal{N}(0,1)$ in the generation step should yield showers resembling the \textsc{Geant4} truth more closely. 
As shown in Fig.~\ref{fig:plots_KLD005vs04}~(bottom left) this improves the CoG-Z distribution compared to the baseline. 
However, there  is a trade-off for other distributions, such as the total energy or energy sum (top center) and the number of hits (top right) which become narrower than the baseline and truth distributions.
This can also be seen in Fig.~\ref{fig:res_mean_KLD005vs04}: 
Except for low energies the energy linearity is better, but the relative width of the energy distributions is on average narrower than the baseline model.

\begin{figure*}[t]
    \centering
    \sidecaption
    \includegraphics[width=0.35\textwidth]{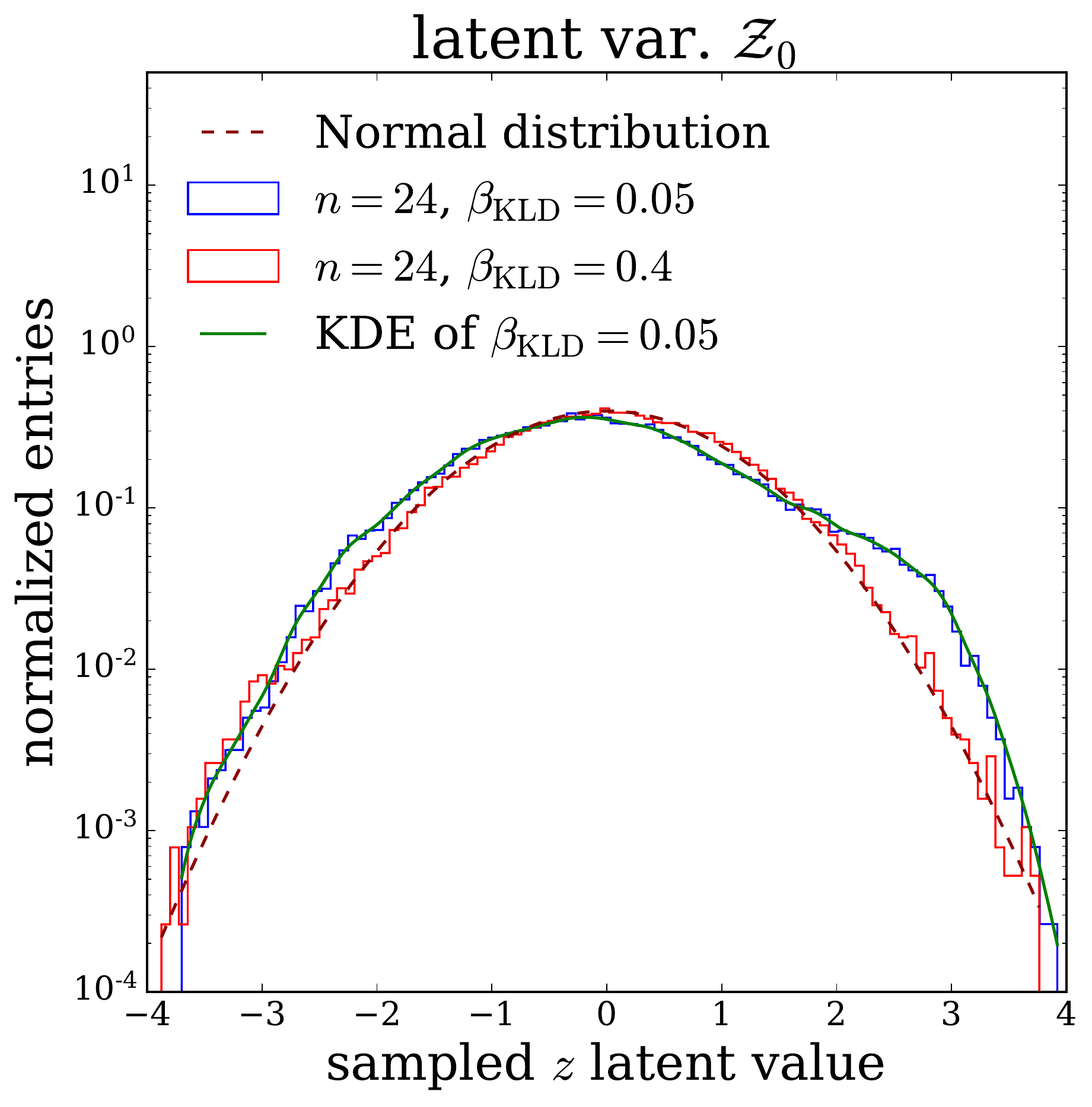}
    \includegraphics[width=0.35\textwidth]{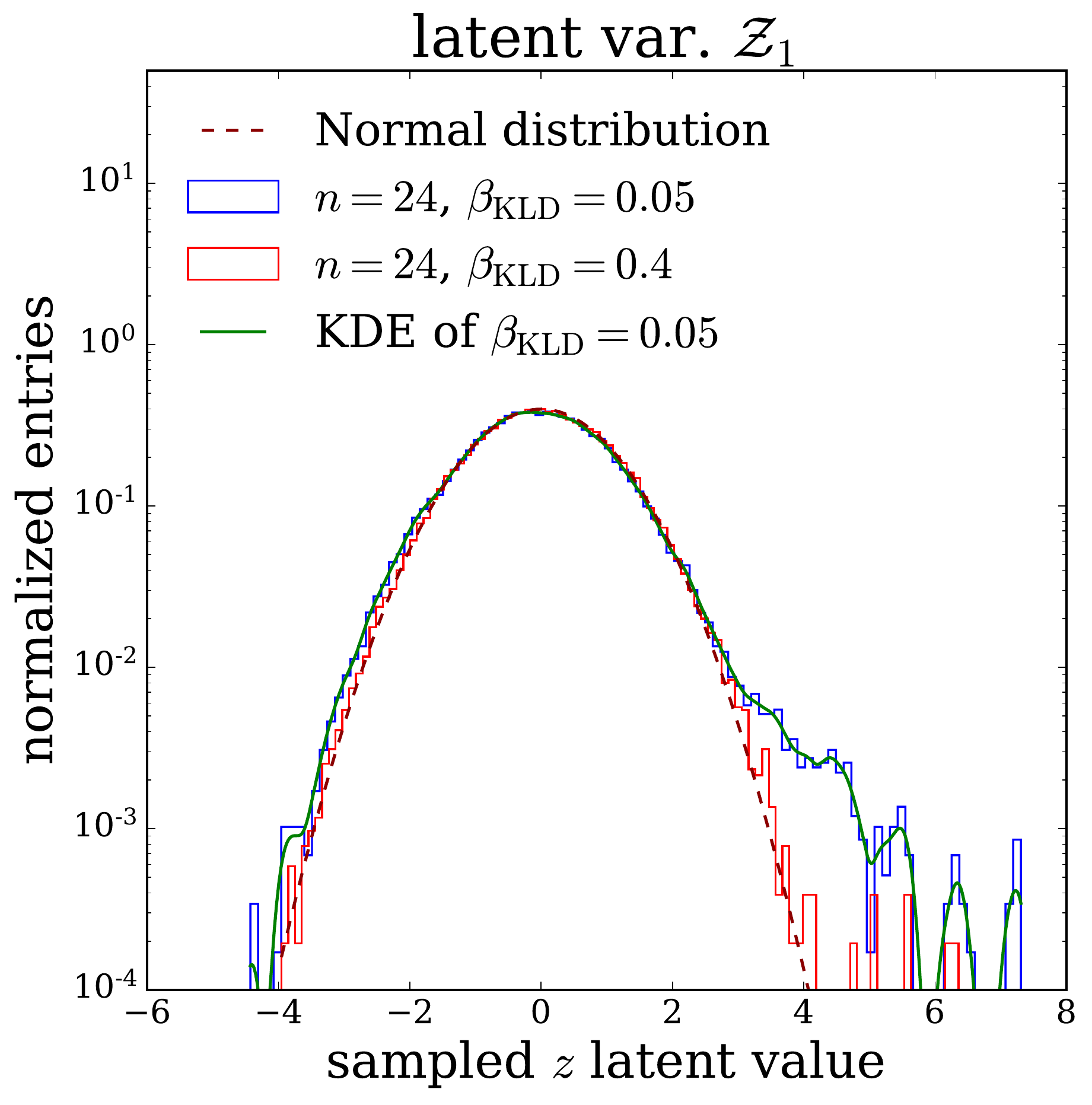}
    \caption{Sampled $z$ values of the highest (left) and second-highest (right) KLD latent variables for 50k shower images for models with a latent size of 24 and $\beta_{KLD} = 0.05$ or $\beta_{KLD} = 0.4$. 
    For reference added lines for a Normal distributions and the Kernel Density Estimate of the $\beta_{KLD}=0.05$ histograms. 
    }
    \label{fig:latent_hist_KDE_comp_varSortNo}
\end{figure*}

Figure~\ref{fig:latent_hist_KDE_comp_varSortNo} illustrates that for 
the highest (left) and second-highest (right)
KLD latent variables, the sampled $z$ distributions for $\beta_{KLD} = 0.4$ are very similar to Normal distributions while they deviate significantly for the baseline value of $\beta_{KLD}=0.05$. 
Although improving the CoG-Z distribution, the overall fidelity score
given in Table~\ref{tab:fidelity_scores_24} 
is slightly worse for $\beta_{KLD} = 0.4$.


\subsection{Sampling from a Kernel Density Estimate}

Another way to improve the generative performance, particularly the CoG-Z distribution, is to utilize latent variables highly correlated to the CoG-Z distribution. Using exactly the same model as in Ref.~\cite{Getting_High} ($\beta_{KLD}=0.05$ , $n=24$) without retraining one can see in Fig.~\ref{fig:latent_hist_KDE_comp_varSortNo} that the encoded distribution deviates from a Standard Normal distribution. In the usual VAE-like setup one would regardless sample these variable from $\mathcal{N}(0,1)$ to generate new samples, thereby ignoring the correlations between the latent space and the shower physics.
Instead one could sample those latent variables from the distribution of the sampled $z_i$ values, which are sample from the encoded $\mathcal{N}(\mu_i, \sigma_i^2)$ distributions. Since at least two variables as well as the incident energy are correlated to the CoG-Z distribution, one needs to account for correlations between latent variables when sampling.
This can be done by encoding a sufficiently large number of showers (i.e. 500k) from the training set, applying a density estimation method such KDE,
and then sampling new latent variables from it. In the BIB-AE case with 24 encoded latent variables plus energy conditioning, this leads to training a KDE of a 25-dimensional space. The resulting KDE kernel can be used as a probability density function for sampling the latent $z$ variables for improved shower generation. 

As shown in Fig.~\ref{fig:plots_KLD005vs04} this KDE sampling approach yields global differential distributions very similar to the \textsc{Geant4} truth; superior results for the CoG-Z distribution and the number of hits distributions in comparison to the other two models.
The linearity in Fig.~\ref{fig:res_mean_KLD005vs04} closely resembles the baseline model, however the relative width of the energy distributions is on average slightly overestimated.
The fidelity score in Table~\ref{tab:fidelity_scores_24} is the best of all tested model configurations. The score for the model of $\beta_{\textrm{KLD}}=0.05$ was chosen as the best out of seven training runs and the same model was used to simply add the KDE sampling step.
This illustrates  another benefit of the KDE approach:  It can be applied to any already trained VAE-like model without expensive re-training.

\section{Summary \& conclusions}
\label{sec:summary}

Improving the simulation of calorimeter showers with generative models is an active topic of research motivated by these tasks' large resource consumption. 
As such generative models still require substantial training efforts and preclude large hyperparameter scans for optimization, we investigate how a better understanding of the latent space can be used to increase performance.
While a BIB-AE architecture was used for these studies, the developed strategies should readily transfer to other generative models with an encoded latent space (i.e. VAE-like but not GAN-like architectures).


We first quantify the information encoded in the latent space and note that for a fixed value of $\beta_{KLD}=0.05$, it saturates at $\approx 45$~nats. However, generative performance --- as measured by a metric defined to take the relevant physical distributions into account --- 
achieves its best value at a latent space of $n=24$ with $\approx 28$~nats.
Put differently, more information encoded in the latent space will not necessarily translate into better generative performance. 

This observation offers an interesting parallel to the information bottleneck principle~\cite{BIB-AE,tishby2000information}. It proposes that for a supervised classification task, the latent space $\mathcal{Z}$ should maximise its mutual information $I$ with the true class labels $\mathcal{C}$  but minimise information irrelevant for classification between  data examples $\mathcal{X}$ and latent space: 
\begin{equation}
  \mathcal{L}_S(\phi) = I_{\phi}(\mathcal{X};\mathcal{Z}) - \beta I(\mathcal{Z};\mathcal{C}).    
\end{equation}
Here $\mathcal{L}_S$ is the supervised optimisation target, we minimise over parametric mappings $\phi$ from data to latent space, and the Lagrange multiplier $\beta$ denotes the trade-off between the two goals.

For unsupervised tasks, no class labels are available, and the problem becomes:
\begin{equation}
  \mathcal{L}_U(\phi) = I_{\phi}(\mathcal{X};\mathcal{Z}) - \beta I(\mathcal{Z};\mathcal{X})    
\end{equation}
which is also the core of the BIB-AE loss formulation~\cite{BIB-AE}. It is a much more challenging compression problem as the entropy of a small number of class labels will, in general, be much smaller --- and therefore easier to encode --- than the entropy of the data distribution.
We observed that without additional constraints, such as restricting the latent space size $n$, more information than needed for good generative performance is encoded in the latent space, suggesting the need for additional regularising constraints.
An interesting open question for future research is therefore how the \textit{useful} encoded information might be quantified. 


Regardless of the model configuration, only a few latent variables of the BIB-AE contain most of the shower information. 
Correlating the latent variables with various shower physics metrics reveals that the center of gravity in z-direction is always encoded into the two highest KLD latent variables.
This encoding can be leveraged for targeted shower generation of photon showers with a specific shower start by sampling from a subspace of the highest KLD variable. 

Furthermore, this observations can help improve the generative fidelity of the BIB-AE model. 
This can be achieved either by lowering the encoded KLD or by sampling directly from the encoded latent space density distribution, e.g. learned via Kernel Density Estimation.
Forcing the latent distributions closer to unit Normal naturally improves physical observables most strongly correlated with the corresponding latent space variables with the highest-KLD values,
and decreases the performance of the others.
The latter approach yields the best results with the additional benefit of applying to the already previously trained BIB-AE model (or any other VAE-like model).

The increasing use of generative machine learning models  motivates a closer look into their learned encoding.
Especially in particle physics, the needed precision for many differential distributions over many orders of magnitude offers a rich laboratory to study the connection between generation fidelity and latent space.
On the one hand, this offers several methods to probe and improve generative performance, for example by identifying poorly modeled distributions for which a discrepancy between encoded-into and sampled-from latent space exists. Resolving this discrepancy yields better-generated showers.
On the other hand, the observed difference between maximum-information 
and best-performance latent space capacity raises an interesting problem for future studies.

\begin{acknowledgement}
\section*{\acknowledgementname}
We would like to thank the Maxwell and National Analysis Facility (NAF) computing
centers at DESY for the smooth operation and technical support.
E. Buhmann is funded by a scholarship of the Friedrich Naumann Foundation for Freedom and by the German Federal Ministry of Science and Research (BMBF) via 
\textit{Verbundprojekts 05H2018 - R\&D COMPUTING
(Pilot\-maß\-nah\-me ErUM-Data) Innovative Digitale
Technologien f\"ur die Erforschung von Universum und
Materie}.
S. Diefenbacher is funded by the Deutsche Forschungsgemeinschaft (DFG, German Re\-search Foundation) 
under Germany’s Excellence Strategy – EXC 2121  ``Quantum Universe" – 390833306. 
E. Eren was funded through the Helmholtz Innovation Pool project AMALEA that provided a stimulating scientific environment for parts of the research done here.
\end{acknowledgement}

\bibliography{1_BIB.bib}
%
%
%
%


\appendix

\section{Fidelity score}
\label{app:fidelity_metric}


Comparing histograms of shower variables such as total energy, number of hits, shower profile and center of gravity as shown in Fig.~\ref{fig:plots_KLD005vs04} is a way to determine the generation performance of the generative model in comparison to the \textsc{Geant4} simulation.
It is however difficult to quantify the model improvement by manually observing these plots.
A quantification of the 'generation performance' or 'fidelity' can be calculated via the difference between the histograms of generated and \textsc{Geant4} observables. This can be done for example by calculating the Jensen-Shannon distance (JSD) by considering each histogram as a discrete probability density distribution.
As an alternative we have calculated a fidelity score based on the area difference between the histograms. This score was comparable to our fidelity score $S_{\text{JSD}}$.
A similar fidelity metric was calculated in Ref.~\cite{Buffer_VAE}.


The JSD can be calculated for each of the six histograms in Fig.~\ref{fig:plots_KLD005vs04}. To have one score combining all six histograms one needs to weight each individual histograms' JSD in comparison to all other JSDs of the same model.
This weighting is done in the following way:
\begin{enumerate}
    \item Calculate JSD for each of the six plots for each model configuration and epoch: $\text{JSD}_{i,m,e}$ with $i$ for 1 in 6 plots, $m$ for 1 in x models, and $e$ for 1 in y epochs that are compared in the score
    \item Calculate the 6 weighting factor for the JSD of each $i$ plot:\\
    $<\text{JSD}_{i}> = \overline{\text{JDS}_{i,m,e}}$ for each plot $i$ 
    \item Calculate the fidelity score $S_{\text{JSD}}$ for each model $m$ and epoch $e$:\\
    $S_{\text{JSD},m,e} = <\text{JSD}_{m,e}> = \frac{1}{6}\sum_i \textrm{JSD}_{i,m,e} \cdot \frac{1}{<\textrm{JSD}_i>}$
\end{enumerate}

\begin{figure*}[t]
    \centering
    \includegraphics[width=0.4\textwidth]{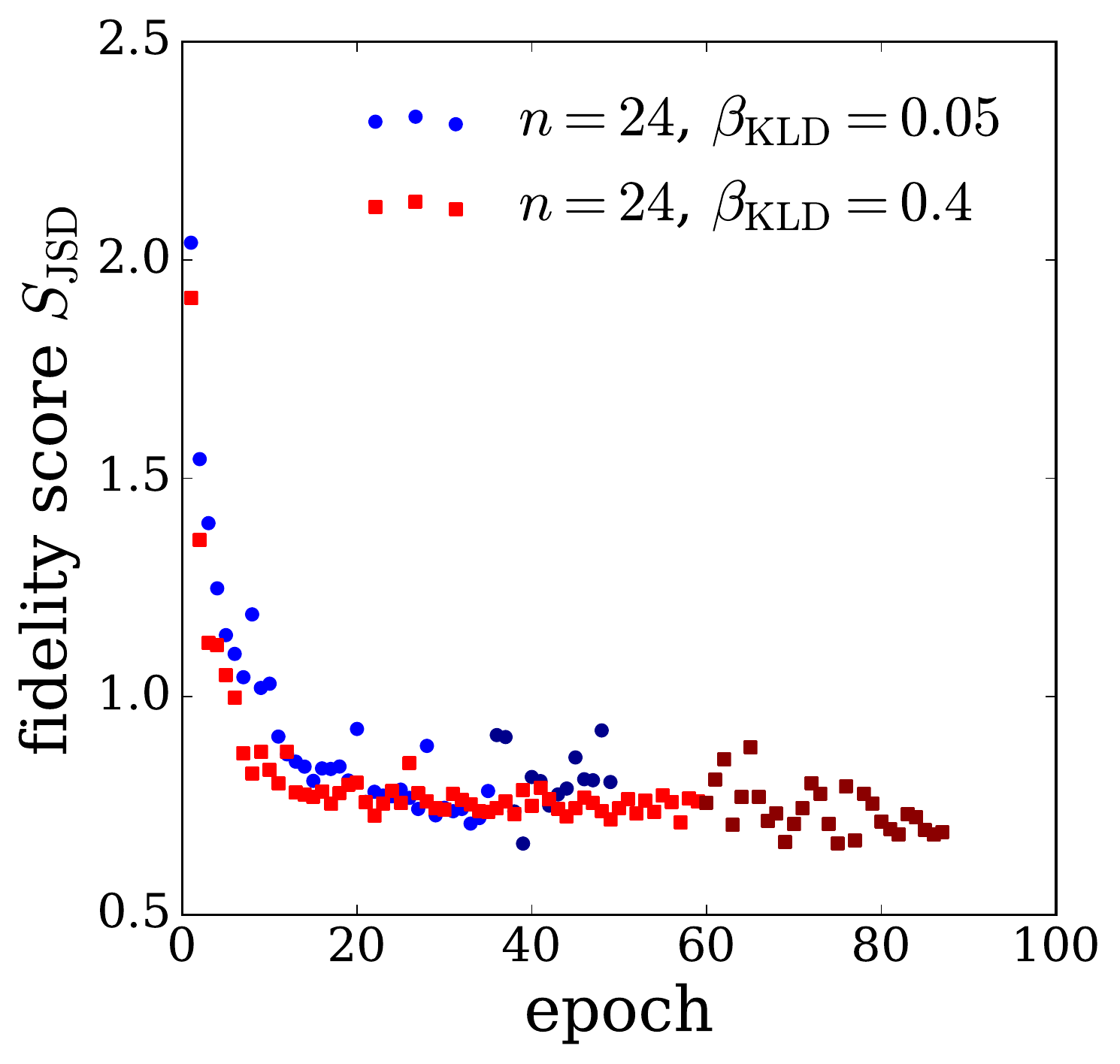}
    \includegraphics[width=0.4\textwidth]{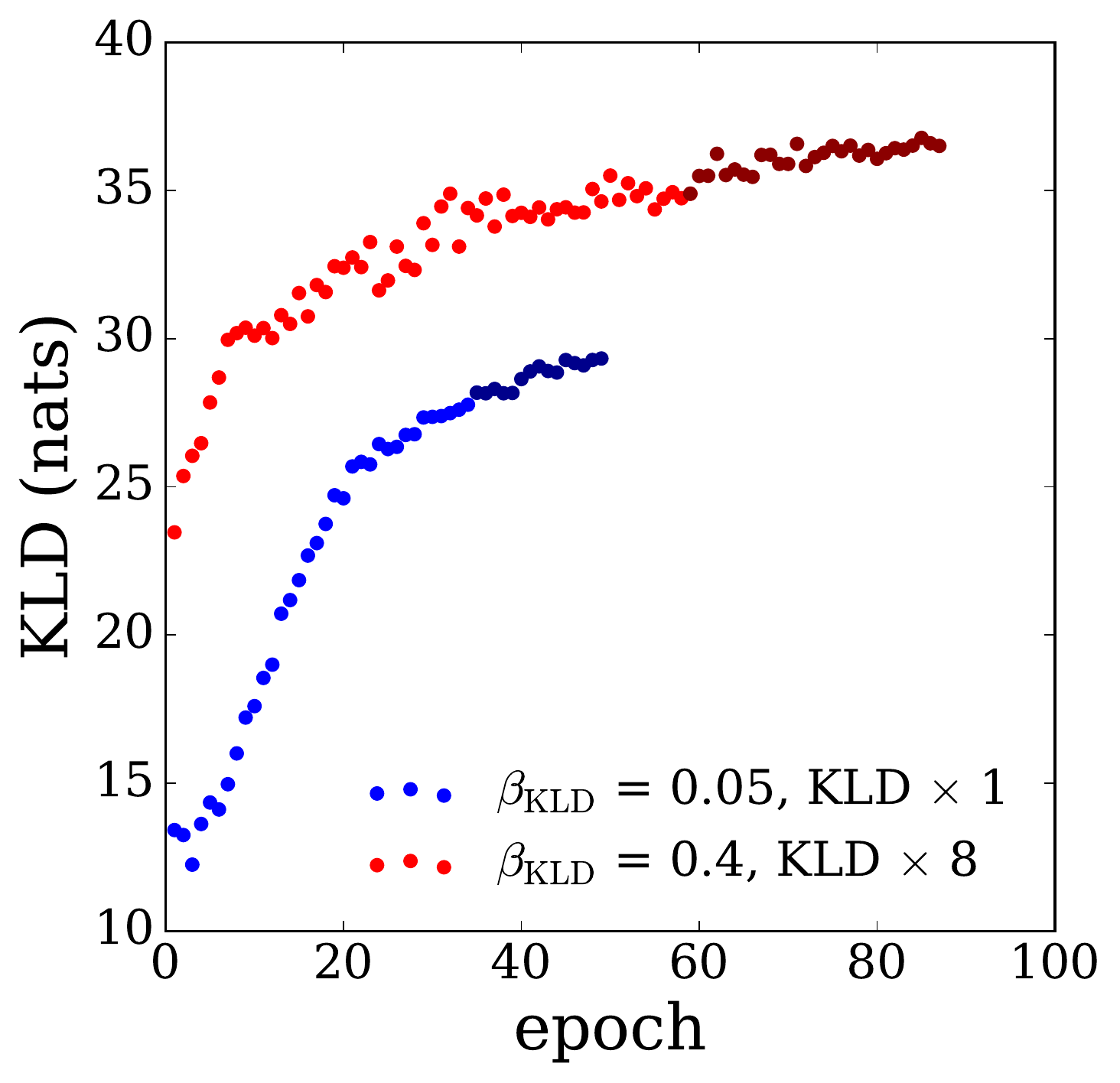}
    \caption{Evolution of the fidelity score $S_{\text{JSD}}$ and the KL divergence over the course the of training for the two models with $\beta_{KLD} = 0.05$ and $\beta_{KLD} = 0.4$.
    Based on the fidelity score the best epochs were chosen (epoch 39 and epoch 87 respectively).
    Color brightness implies training with or without the Post-Processor network (see Sec.~\ref{sec:model}).
    }
    \label{fig:JSD_AD_KLD_example}
\end{figure*}

An example of this weighted $S_{\text{JSD}}$ score is shown in Fig.~\ref{fig:JSD_AD_KLD_example} for an epoch-wise scan during the training of two models with different $\beta_{KLD}$ weights; each with and without the Post-Processor network. Note that the KLD is increasing with each epoch and saturates over time. However, a higher KLD does not necessarily correlate with a lower fidelity score. 

\end{document}